\documentclass{article}

\usepackage{microtype}
\usepackage{graphicx}
\usepackage{subfigure}
\usepackage{booktabs} 
\usepackage{amssymb}
\usepackage[cmex10]{amsmath}
\newtheorem{lemma}{\bf Lemma}
\newtheorem{theorem}{\bf Theorem}
\usepackage{hyperref}

\usepackage{algorithm}
\usepackage{algorithmic}

\usepackage{color}
\definecolor{myc1}{rgb}{0,0,0}

\usepackage[accepted]{icml2020}

\icmltitlerunning{Delay Minimization for Federated Learning Over Wireless Communication Networks}

\begin{document}

\twocolumn[
\icmltitle{Delay Minimization for Federated Learning Over \\ Wireless Communication Networks}



\icmlsetsymbol{equal}{*}

\begin{icmlauthorlist}
\icmlauthor{Zhaohui Yang}{equal,kcl}
\icmlauthor{Mingzhe Chen}{equal,pri,hk}
\icmlauthor{Walid Saad}{vt}
\icmlauthor{Choong Seon Hong}{korea}
\icmlauthor{Mohammad Shikh-Bahaei}{kcl}
\icmlauthor{H. Vincent Poor}{pri}
\icmlauthor{Shuguang Cui}{hk}
\end{icmlauthorlist}

\icmlaffiliation{kcl}{Centre for Telecommunications Research, Department of Engineering, King's College London, UK}
\icmlaffiliation{pri}{Electrical Engineering Department, Princeton University, USA}
\icmlaffiliation{hk}{Shenzhen Research Institute of Big Data and School of Science and Engineering, the Chinese University of Hong Kong, China}
\icmlaffiliation{vt}{Wireless@VT, Bradley Department of Electrical and Computer Engineering,  Virginia Tech, USA}
\icmlaffiliation{korea}{Department of Computer Science and Engineering, Kyung Hee University, Rep. of Korea}

\icmlcorrespondingauthor{Zhaohui Yang}{yang.zhaohui@kcl.ac.uk}
\icmlcorrespondingauthor{Mingzhe Chen}{mingzhec@princeton.edu}

\icmlkeywords{Machine Learning, ICML}
]



\printAffiliationsAndNotice{\icmlEqualContribution} 

\begin{abstract}
In this paper, the problem of delay minimization for federated learning (FL) over wireless communication networks is investigated. In the considered model, each user exploits limited local computational resources to train a local FL  model with its collected data and, then, sends the trained FL model parameters to a base station (BS) which aggregates the local FL models and broadcasts the aggregated FL model back to all the users. Since FL involves learning model exchanges between the users and the BS, both computation and communication latencies are determined by the required learning accuracy level, which affects the convergence rate of the FL algorithm. This joint learning and communication problem is formulated as a delay minimization problem, where it is proved that the objective function is a convex function of the learning accuracy. Then, a bisection search algorithm is proposed to obtain the optimal solution. Simulation results show that the proposed algorithm can reduce delay by up to 27.3\% compared to conventional FL methods.
\end{abstract}

\vspace{-2.5em}
\section{Introduction}
\vspace{-.5em}
In future wireless systems, due to privacy constraints and limited communication resources for data transmission, it is impractical for all wireless devices to transmit all of their collected data to a data center that can  implement centralized machine learning algorithms for data analysis \cite{wang2018edge,8755300,9110869,8752012,9031435}.
To this end, distributed edge learning approaches, such as federated learning (FL), were proposed \cite{saad2019vision,Wireless2018Park,chen2018federated,samarakoon2018distributed,8839651,9013160}.
In FL, the wireless devices individually establish local learning models and cooperatively build a global learning model by uploading the local learning model parameters to a base station (BS) instead of sharing training data\cite{mcmahan2016communication,8851249,8664630}.
To implement FL over wireless networks, the wireless devices must transmit their local training results over wireless links \cite{zhu2018towards}, which can affect the FL performance, because both local training and wireless transmission introduce delay.
Hence, it is necessary to optimize the delay for wireless FL implementation.



Some of the challenges of FL over wireless networks have been studied in \cite{zhu2018low,ahn2019wireless,
Yang2018FLOA,Zeng2019EEFL,chen2019joint,tran2019federated}.
To minimize latency, a broadband analog aggregation multi-access scheme for FL was designed in \cite{zhu2018low}.
The authors in \cite{ahn2019wireless} proposed an FL  implementation scheme between devices and access point  over Gaussian multiple-access channels.
To improve the statistical learning performance for on-device distributed training, the authors in \cite{Yang2018FLOA} developed a sparse and low-rank modeling approach.
The work in in \cite{Zeng2019EEFL} proposed
an energy-efficient strategy for bandwidth allocation
 with the goal of reducing devices'
sum energy consumption while meeting the required learning performance.
 However, the prior works \cite{konevcny2016federated,zhu2018low,ahn2019wireless,
Yang2018FLOA,Zeng2019EEFL} focused on the delay/energy consumption for wireless consumption without considering the delay/energy tradeoff between learning and transmission.
Recently, in \cite{chen2019joint} and \cite{tran2019federated}, the authors considered both local learning and wireless transmission energy.
In \cite{chen2019joint}, the authors investigated the FL loss function minimization problem with taking into account packet errors over wireless links.
However, this prior work ignored the computation delay of local FL model.
The authors in \cite{tran2019federated} considered the sum learning and transmission energy minimization problem for FL,
where all users transmit learning results to the BS.
However, the solution in \cite{tran2019federated} requires all users to upload their learning model synchronously.
The main contribution of this paper is a framework for optimizing FL over wireless networks. In particular, we consider a wireless-powered FL algorithm in which each user locally computes its FL model parameters under a given learning accuracy and the BS  broadcasts the aggregated FL model parameters to all users. Considering the tradeoff between local computation delay and wireless transmission delay,
we formulate a joint transmission and computation optimization problem aiming to minimize the delay for FL.
We theoretically
show that the delay is a convex function of the learning accuracy. 
Based on the theoretical finding, we propose a bisection-based algorithm to obtain the optimal solution.
\vspace{-1em}
\section{System Model and Problem Formulation}
\vspace{-.5em}
%

\begin{figure}
\centering
\includegraphics[width=3.0in]{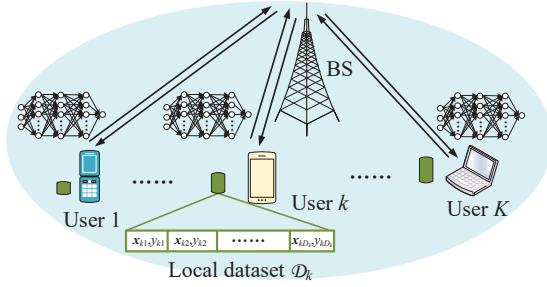}
\vspace{-1.5em}
\caption{FL over wireless communication networks.}
\vspace{-1.5em}
\label{sys}
\end{figure}

Consider a cellular network that consists of one BS serving a set $\mathcal K$ of $K$ users, as shown in Fig.~\ref{sys}.
Each user $k$ has a local dataset $\mathcal D_k$ with $D_k$ data samples.
For each dataset $\mathcal D_k=\{\boldsymbol x_{kl},y_{kl}\}_{l=1}^{D_k}$, $\boldsymbol x_{kl}\in\mathbb R^d$ is an input vector of user $k$ and $y_{kl}$ is its corresponding output\footnote{For simplicity, this paper only considers
an FL algorithm with a single output. Our approach can be extended to the
case with multiple outputs \cite{konevcny2016federated}.}.
\vspace{-0.5em}
\subsection{FL Model}
\vspace{-0.5em}
For FL, we define a vector $\boldsymbol w$ to capture the parameters related to the global FL  model that  is  trained by all datasets. Hereinafter, the FL model that is trained by all users' data set is called \emph{global FL model}, while the FL model that is trained by each user's dataset is called \emph{local FL model}.
We introduce the loss function $f(\boldsymbol w,\boldsymbol x_{kl}, y_{kl})$, that captures the  FL performance over input vector $\boldsymbol{x}_{kl}$ and output $y_{kl}$.
For different learning tasks, the loss function will be different.
Since the dataset of user $k$ is $\mathcal D_k$, the total loss function of user $k$ will be:
\vspace{-1em}
\begin{equation}\label{sys0eq1}\vspace{-1em}
F_k(\boldsymbol w)=\frac 1 {D_k} \sum_{l=1}^{D_k} f(\boldsymbol w,\boldsymbol x_{kl}, y_{kl}).
\end{equation}


In order to deploy FL, it is necessary to train the underlying model. Training is done in order to compute the global FL model for all users without sharing their local datasets due to privacy and communication issue.
The FL training problem can be formulated as follows \cite{wang2018edge}:
\vspace{-0.5em}
\begin{equation}\label{sys0eq2}\vspace{-0.5em}
\min_{\boldsymbol w} F(\boldsymbol w)=\sum_{k=1}^K \frac{D_k}{D} F_k(\boldsymbol w)=\frac{1}{D}\sum_{k=1}^K\sum_{l=1}^{D_k} f(\boldsymbol w,\boldsymbol x_{kl}, y_{kl}),
\end{equation}
where $D=\sum_{k=1}^KD_k$ is the total data samples of all users.

To solve problem \eqref{sys0eq2}, we adopt the FL algorithm in \cite{konevcny2016federated}, which is summarized in Algorithm~1.
\begin{algorithm}[t]
\caption{FL Algorithm}
\scriptsize
\begin{algorithmic}[1]
 \STATE Initialize 
 global regression vector $\boldsymbol w^0$ and iteration number $n=0$.
 \REPEAT
 \STATE Each user $k$ computes $\nabla  F_k(\boldsymbol w^{(n)})$ and sends it to the BS.
 \STATE The BS computes $\nabla  F(\boldsymbol w^{(n)})= \frac 1 K\sum_{k=1}^K \nabla  F_k(\boldsymbol w^{(n)})$,
 which is broadcast to all users.
 \STATE  \textbf{parallel for} user $k\in\mathcal K$
\STATE \quad  Solve local FL problem  \eqref{sys0eq3_1} with a given learning accuracy $\eta$ and the solution is $\boldsymbol h_k^{(n)}$.
 \STATE \quad Each user sends $\boldsymbol h_k^{(n)}$ to the BS.
  \STATE \textbf{end for}
  \STATE   The BS computes $\boldsymbol w^{(n+1)}=\boldsymbol w^{(n)}+\frac {1} K \sum_{k=1}^K\boldsymbol h_k^{(n)}$ and broadcasts the value to all users.
 \STATE Set $n=n+1$.
 \UNTIL the accuracy $\epsilon_0$ of problem \eqref{sys0eq2} is obtained.
\end{algorithmic}
\end{algorithm}
In Algorithm 1, at each iteration of the FL algorithm, each user downloads the global FL model parameters from the BS for local computing, while  the BS periodically gathers the local FL model parameters from all users and sends the updated global FL model parameters back to all users.
We define $\boldsymbol w^{(n)}$ as the global FL parameter at a given iteration $n$.
Each user computes the local FL problem:
\vspace{-.5em}
\begin{align}\label{sys0eq3_1}\vspace{-.5em}
  \min_{\boldsymbol h_k\in\mathbb R^d}\quad G_k&(\boldsymbol w^{(n)}, \boldsymbol h_k)\triangleq  F_k(\boldsymbol w^{(n)}+ \boldsymbol h_k)
  \nonumber \\&
-(\nabla  F_k(\boldsymbol w^{(n)})-  \xi \nabla  F(\boldsymbol w^{(n)}))^T \boldsymbol h_k,
\end{align}
by using the gradient method with a given accuracy. 
 In problem \eqref{sys0eq3_1}, $\xi$ is a constant value.
The solution $\boldsymbol h_k$ in problem \eqref{sys0eq3_1} means the {\color{myc1}{updated value}} of local FL parameter for user $k$ in each iteration, 
i.e., $\boldsymbol w^{(n)}+\boldsymbol h_k$ denotes user $k$' local FL parameter at the $n$-th iteration.
Since it is hard to obtain the actual optimal solution of problem \eqref{sys0eq3_1}, we obtain a solution of \eqref{sys0eq3_1} with some accuracy.
The solution $\boldsymbol h_k^{(n)}$ of problem \eqref{sys0eq3_1} at the $n$-th iteration with accuracy $\eta$  means that
\vspace{-1em}
   \begin{align}\label{sys0eq3_6}\vspace{-1em}
G_k &(\boldsymbol w^{(n)}, \boldsymbol h_k^{(n)})-G_k(\boldsymbol w^{(n)}, \boldsymbol h_k^{(n)*})
\nonumber\\
&
\leq  \eta (G_k (\boldsymbol w^{(n)}, \boldsymbol 0)-G_k(\boldsymbol w^{(n)}, \boldsymbol h_k^{(n)*})),
 \end{align}
where $\boldsymbol h_k^{(n)*}$ is the actual optimal solution of problem \eqref{sys0eq3_1}.



%


In Algorithm 1, the iterative method involves
a number of global iterations (i.e., the value of $n$ in Algorithm 1) to achieve a global accuracy $\epsilon_0$ of global FL model.
The solution $\boldsymbol w^{(n)}$ of problem \eqref{sys0eq2} with accuracy $\epsilon_0$  means that
\vspace{-.5em}
   \begin{align}\label{sys0eq3_7}\vspace{-1em}
F  (\boldsymbol w^{(n)})-F(\boldsymbol w^*)\leq \epsilon_0 (F  (\boldsymbol w^{(0)})-F(\boldsymbol w^*)),
 \end{align}
where $\boldsymbol w^{*}$ is the actual optimal solution of problem \eqref{sys0eq2}.

To analyze the convergence of Algorithm~1, we assume that  $ F_k (\boldsymbol{w} )$ is $L$-Lipschitz continuous and $\gamma$-strongly convex, i.e.,
\vspace{-0.5em}
\begin{equation}\label{flconrproofeq0}\vspace{-0.5em}
\gamma\boldsymbol{I} \preceq
\nabla^2 F_k (\boldsymbol{w} ) \preceq L \boldsymbol{I}, \quad \forall k\in \mathcal K.
\end{equation}
Under assumption \eqref{flconrproofeq0}, we provide the following lemma about convergence rate of  Algorithm~1.

\vspace{-1em}
\begin{lemma}
If we run Algorithm 1 with $0<\xi\leq \frac{\gamma}{L}$ for
\vspace{-0.5em}
\begin{equation}\label{sys1eq5_1}\vspace{-.5em}
n\geq\frac{a}{1-\eta}\triangleq I_0,
\end{equation}
iterations with
$a=\frac{2L^2
}{\gamma^2\xi}\ln\frac{1}{\epsilon_0}$,
we have $F  (\boldsymbol w^{(n)})-F(\boldsymbol w^*)\leq \epsilon_0 (F  (\boldsymbol w^{(0)})-F(\boldsymbol w^*))$.
\end{lemma}

The proof of Lemma 1 can be found in \cite{yang2019energy}.
From Lemma 1, we can find that the number of global iterations $n$ increases with the local accuracy.
This is because more iterations are needed if the local computation has a low accuracy.
\vspace{-.5em}
\subsection{Computation and Transmission Model}
\vspace{-.5em}
The FL procedure between the users and their serving BS consists of three steps in each  iteration: Local computation at each user (using several local iterations), local FL parameter transmission for each user, and result aggregation and broadcast at the BS.
During the local computation step, each user calculates its local FL parameters by using its local dataset and the received global FL parameters.
\vspace{-.5em}
\subsubsection{Local Computation}
\vspace{-.5em}
We solve the local learning problem \eqref{sys0eq3_1} by using the gradient method.
In particular, the gradient procedure in the $(i+1)$-th iteration is given by:
\vspace{-.5em}
\begin{equation}\label{applemmacon2eq1}\vspace{-.5em}
\boldsymbol h_k^{(n),(i+1)}=\boldsymbol h_k^{(n),(i)}- \delta\nabla G_k(\boldsymbol w^{(n)}, \boldsymbol h_k^{(n),(i)}),
\end{equation}
where $\delta$ is the step size, $\boldsymbol h_k^{(n),(i)}$ is the value of $\boldsymbol h_k$
at the $i$-th local iteration with given vector $\boldsymbol w^{(n)}$, and $\nabla G_k(\boldsymbol w^{(n)}, \boldsymbol h_k^{(n),(i)})$ is the gradient of function $G_k(\boldsymbol w^{(n)}, \boldsymbol h_k )$ at point $\boldsymbol h_k=\boldsymbol h_k^{(n),(i)}$.
We set the initial solution $\boldsymbol h_k^{(n),(0)}=\boldsymbol 0$.

Next, we provide the number of local iterations needed to achieve a local accuracy $\eta$ in \eqref{sys0eq3_6}.
We set $v=\frac{2}{(2-L\delta)\delta\gamma}$.

\vspace{-1em}
\begin{lemma}\label{lemmacon2}
If we set step $\delta< \frac{2}{L}$ and run the gradient method for
$
i\geq v\log_2(1/\eta) 
$
iterations 
at each user,
we can solve local FL problem  \eqref{sys0eq3_1} with an accuracy $\eta$.
\end{lemma}

 The proof of Lemma 2 can be found in paper \cite{yang2019energy}.
Let $f_k$ be the computation capacity of user $k$, which is measured by the number of CPU cycles per second.
The computation time at user $k$ needed for data processing is:
\vspace{-0.5em}
\begin{equation}\label{sys1eq5_2}\vspace{-0.5em}
\tau_{k} =\frac{v C_k D_k \log_2(1/\eta)}{f_k} =\frac{A_k \log_2(1/\eta)}{f_k}, \quad \forall k\in\mathcal K,
\end{equation}
where $C_{k}$ (cycles/bit) is the number of CPU cycles required for computing one sample data at user $k$, $v\log_2(1/\eta)$ is the number of local iterations for each user as given by Lemma~2, and $A_k=vC_k D_k$.

\vspace{-.5em}
\subsubsection{Wireless Transmission}\vspace{-.5em}
After local computation, all users upload their local FL parameters to the BS via frequency domain multiple access (FDMA).
The achievable rate of user $k$ can be given by:
\vspace{-1em}
\begin{equation}\label{sys1eq7}\vspace{-.5em}
r_k=b_k\log_2\left(1+ \frac{g_kp_k}{N_0b_k}
\right), \quad \forall k\in\mathcal K,
\end{equation}
where $b_k$ is the bandwidth allocated to user $k$, $p_k$ is the transmit power of user $k$, $g_k$ is the channel gain between user $k$ and the BS, and $N_0$ is the power spectral density of the Gaussian noise.
Due to the limited bandwidth, we have $
\sum_{k=1}^K b_k \leq B,
$
where $B$ is the total bandwidth.

In this step, user $k$ needs to upload the local FL parameters 
 to the BS.
Since the dimensions of the vector $\boldsymbol h_k^{(n)}$ are fixed for all users, the data size that each user needs to upload is constant, and can be denoted by $s$.
To upload data of size $s$ within  transmit time $t_k$, we must have:
$
t_k{r_k} \geq s.
$

\vspace{-.5em}
\subsubsection{Information Broadcast}\vspace{-.5em}
In this step, the BS aggregates the global prediction model parameters.
The BS broadcasts the global prediction model parameters to all users in the downlink.
Due to the high power of the BS and large downlink bandwidth, we ignore the downlink time.
Note that the local data $\mathcal D_k$ is not accessed by the BS, so as to protect the privacy of users, as is required by FL.
%
The delay of each user includes the local computation time and transmit time.
Based on \eqref{sys1eq5_1} and \eqref{sys1eq5_2}, the delay $T_k$ of user $k$ will be:
\vspace{-1em}
\begin{align}\vspace{-1em}
T_k=I_0(\tau_k+t_k)
=\frac a {1-\eta}\left(\frac{A_k\log_2(1/\eta)}{f_k}+t_k
\right).
\end{align}
We define $T=\max_{k\in\mathcal K} T_k$ as the delay for training the whole FL algorithm.

\vspace{-.5em}
\subsection{Problem Formulation}
\vspace{-.5em}
 We now pose the delay minimization problem:
\begin{subequations}\label{sys1min1}\vspace{-1em}
\begin{align}
\mathop{\min}_{T, \boldsymbol t, \boldsymbol b, \boldsymbol f, \boldsymbol p, \eta} \:&
T
\tag{\theequation}  \\
\textrm{s.t.} \quad\:\:\:
&\frac a {1-\eta}\left(  \frac{A_k\log_2(1/\eta)}{f_k} +t_k
\right)\leq T,\quad \forall k\in \mathcal K,\\
& t_kb_k\log_2\left(1+ \frac{g_kp_k}{N_0b_k}
\right) \geq s, \quad \forall k \in \mathcal K,\\
& \sum_{k=1}^K b_k\leq B, \\
& 0\!\leq \!f_k\!\leq\! f_k^{\max},0\!\leq \!p_k\!\leq\! p_k^{\max}, \:\forall k \in \mathcal K,\\
& 0\leq \eta\leq 1,\\
& t_k\geq 0, b_k \geq 0, \quad \forall k \in \mathcal K,
\end{align}
\end{subequations}
where $\boldsymbol t=[t_1, \cdots, t_K]^T$,
$\boldsymbol b=[b_1, \cdots, b_K]^T$,
$\boldsymbol f=[f_1, \cdots, f_K]^T$, and
$\boldsymbol p=[p_1, \cdots, p_K]^T$.
 $f_k^{\max}$ and $p_k^{\max}$ are, respectively, the maximum local computation capacity and maximum transmit power of user $k$.
(\ref{sys1min1}a) indicates that the execution time of the local tasks and the transmit time for all users should not exceed the delay of the whole FL algorithm.
The data transmission constraint is given by (\ref{sys1min1}b),
while the bandwidth constraint is given by (\ref{sys1min1}c).
(\ref{sys1min1}d) represents the maximum local computation capacity and transmit power limits of all users.
The accuracy constraint is given by (\ref{sys1min1}e).




\vspace{-1em}
\section{Optimal Resource Allocation}\vspace{-.5em}

  Although the delay minimization problem (\ref{sys1min1}) is nonconvex due to constraints (\ref{sys1min1}a)-(\ref{sys1min1}b), the globally optimal solution is shown to be obtained by using the bisection method.
\vspace{-.5em}
\subsection{Optimal Resource Allocation}\vspace{-.5em}
Let $(T^*, \boldsymbol t^*, \boldsymbol b^*, \boldsymbol f^*, \boldsymbol p^*, \eta^*)$ be the optimal solution  of problem (\ref{sys1min1}).
We provide the following lemma about the feasibility conditions of problem (\ref{sys1min1}).
{\color{myc1}{
\begin{lemma}\label{lemmatime1}
Problem (\ref{sys1min1}) with fixed $T<T^*$ is always feasible, while problem (\ref{sys1min1}) with fixed $T>T^*$ is infeasible.
\end{lemma}
}}

\itshape {Proof:}  \upshape
Assume that $(\bar T, \bar{ \boldsymbol t}, \bar{\boldsymbol b}, \bar{\boldsymbol f}, \bar{\boldsymbol p},\bar\eta)$ is a feasible solution of problem (\ref{sys1min1}) with  $T=\bar T<T^*$.
Then, solution $(\bar T, \bar{ \boldsymbol t}, \bar{\boldsymbol b}, \bar{\boldsymbol f}, \bar{\boldsymbol p}, \bar\eta)$ is feasible with lower value of the objective function than solution $(T^*, \boldsymbol t^*, \boldsymbol b^*, \boldsymbol f^*, \boldsymbol p^*,\eta^*)$, which contradicts the fact that $(T^*, \boldsymbol t^*, \boldsymbol b^*, \boldsymbol f^*, \boldsymbol p^*,\eta^*)$ is the optimal solution.
For problem (\ref{sys1min1}) with  $T=\bar T>T^*$, we can always construct a feasible solution $(\bar T, \boldsymbol t^*, \boldsymbol b^*, \boldsymbol f^*, \boldsymbol p^*, \eta^* )$ to problem (\ref{sys1min1})  by checking all constraints.
  \hfill $\Box$


{\color{myc1}{According to Lemma 3, we can use the bisection method to obtain the optimal solution of problem \eqref{sys1min1}. 
Denote
\vspace{-.5em}
\begin{equation}\vspace{-.5em}
T_{\min}=0, T_{\max}=\max_{k\in\mathcal K}
 \frac{2aA_k }{f_k^{\max}} +\frac{2aKs}{ B\log_2\left(1+ \frac{g_kp_k^{\max}K}{N_0 B}
\right)}.
\end{equation}
If  $T>T_{\max}$, problem (\ref{sys1min1}) is always feasible by setting
$f_k=f_k^{\max}$, $p_k=p_k^{\max}$, $b_k=\frac{B}{K}$, $\eta=\frac 12$, and
\vspace{-.5em}
\begin{equation}\vspace{-.5em}
t_k=\frac{Ks}{ B\log_2\left(1+ \frac{g_kp_k^{\max}K}{N_0 B}
\right)}. 
\end{equation}
Hence, the optimal $T^*$ of problem (\ref{sys1min1}) must lie in the interval $(T_{\min}, T_{\max})$.
At each step, the bisection method divides the interval in two by computing the midpoint $T_{\text{mid}}=(T_{\min}+T_{\max})/2$. There are now only two possibilities: 1) if problem (\ref{sys1min1}) with $T=T_{\text{mid}}$ is feasible,  we have $T^*\in(T_{\min},T_{\text{mid}}]$ and
2) if problem (\ref{sys1min1}) with $T=T_{\text{mid}}$ is infeasible, we have  $T^*\in(T_{\text{mid}},T_{\max})$.
The bisection method selects the subinterval that is guaranteed to be a bracket as the new interval to be used in the next step. As such an interval that contains the optimal $T^*$ is reduced in width by 50\% at each step. The process continues until the interval is sufficiently small.}}

With a fixed $T$, we still need to check whether there exists a feasible solution satisfying constraints (\ref{sys1min1}a)-(\ref{sys1min1}g).
From constraints (\ref{sys1min1}a) and (\ref{sys1min1}c), we can see that it is always efficient  to utilize the maximum computation capacity, i.e.,
$f_k^*=f_{k}^{\max},   \forall k \in \mathcal K$.
In addition, from (\ref{sys1min1}b) and (\ref{sys1min1}d), we can see that minimizing the delay can be done by having:
$p_k^*=p_{k}^{\max}, \forall k \in \mathcal K$.
Substituting the maximum computation capacity and maximum transmission power into (\ref{sys1min1}), delay minimization problem becomes:
\begin{subequations}\label{time2min2}\vspace{-0.5em}
\begin{align}
 \min_{ T, \boldsymbol t, \boldsymbol b, \eta} \quad\: &T
  \tag{\theequation}  \\
\textrm{s.t.} \quad\:
& t_k
 \leq \frac {(1-\eta)T}a +\frac{A_k\log_2\eta}{f_k^{\max}},\quad \forall k\in \mathcal K,\\
& \frac{s}{t_k} \leq b_k\log_2\left(1+ \frac{g_kp_k^{\max}}{N_0b_k}
\right), \quad \forall k \in \mathcal K,\\
& \sum_{k=1}^K b_k\leq B, \\
& 0\leq \eta\leq 1,\\
& t_k\geq 0, b_k \geq 0, \quad \forall k \in \mathcal K.
\end{align}
\end{subequations}

We provide the sufficient and necessary condition for the feasibility of set (\ref{time2min2}a)-(\ref{time2min2}e)
using the following lemma.

\begin{lemma}\label{lemmatime11}
With a fixed $T$, set (\ref{time2min2}a)-(\ref{time2min2}e) is nonempty if an only if
\vspace{-.5em}
\begin{equation}\label{time2min2eq1}\vspace{-.5em}
 B\geq \min_{0\leq \eta\leq 1} \quad \sum_{k=1}^K u_k(v_k(\eta)),
\end{equation}
where
\vspace{-0.5em}
\begin{equation} \label{time2min2eq2}\vspace{-0.5em}
u_k(\eta)=-\frac{(\ln2)  \eta }
{ W\left(-\frac{(\ln2) N_0 \eta }{g_kp_k^{\max}} \text e^{-\frac{(\ln2) N_0 \eta }{g_kp_k^{\max}}}
\right)+\frac {(\ln2) N_0\eta}{g_kp_k^{\max}}},
\end{equation}
and
\vspace{-1em}
\begin{equation}\label{time2min2eq3}\vspace{-1em}
v_k(\eta)=\frac{s}{\frac {(1-\eta)T}a +\frac{A_k\log_2\eta}{f_k^{\max}}}.
\end{equation}
\end{lemma}

\itshape {Proof:}  \upshape
To prove this, we first define a function $y=x \ln\left( 1+ \frac{1}{x}
\right)$ with $x>0$.
Then, we have
\vspace{-.5em}
\begin{equation}\label{apptime2min2eq5_1}\vspace{-.5em}
 y'=\ln\left( 1+ \frac{1}{x}
\right) -\frac{1}{x+1},
 y''= -\frac{1}{x(x+1)^2}<0.
\end{equation}

According to  \eqref{apptime2min2eq5_1}, $y'$ is a  decreasing function.
Since $\lim_{t_i \rightarrow +\infty} y' =0$, we have $y'>0$ for all $0<x<+\infty$.
Hence, $y$ is an increasing function, i.e., the right hand side of (\ref{time2min2}b) is an increasing function of bandwidth $b_k$.
To ensure that the maximum bandwidth constraint (\ref{time2min2}c) can be satisfied, the left hand side of (\ref{time2min2}b) should be as small as possible, i.e., $t_k$ should be as long as possible.
Based on  (\ref{time2min2}a), the optimal time allocation should be:
\vspace{-.5em}
\begin{equation}\label{apptime2min2eq5_3}\vspace{-.5em}
t_k^*=\frac {(1-\eta)T}a +\frac{A_k\log_2\eta}{f_k^{\max}},\quad \forall k\in \mathcal K.
\end{equation}

Substituting \eqref{apptime2min2eq5_3} into (\ref{time2min2}b), we can construct  the following problem:
\vspace{-.5em}
\begin{subequations}\label{apptime2min5}\vspace{-.5em}
\begin{align}
 \min_{\boldsymbol b, \eta} \: & \sum_{k=1}^K b_k
  \tag{\theequation}  \\
\textrm{s.t.} \:
& v_k(\eta)\leq b_k\log_2\left(1+ \frac{g_kp_k^{\max}}{N_0b_k}
\right),\: \forall k \in \mathcal K,\\
& 0\leq \eta\leq 1,\\
& b_k \geq 0, \quad \forall k \in \mathcal K,
\end{align}
\end{subequations}
where $v_k(\eta)$  is defined in \eqref{time2min2eq3}.
We can observe that set (\ref{time2min2}a)-(\ref{time2min2}e) is nonempty if an only if the optimal objective value of \eqref{apptime2min5} is less than $B$.
Since the right hand side of (\ref{time2min2}b) is an increasing function,
(\ref{time2min2}b) should hold with equality for the optimal solution of problem \eqref{apptime2min5}.
Setting  (\ref{time2min2}b) with equality, problem \eqref{apptime2min5} reduces to \eqref{time2min2eq1}. \hfill $\Box$

To effectively solve \eqref{time2min2eq1} in Lemma \ref{lemmatime11}, we provide the following lemma.
\begin{lemma}\label{lemmatime2}
In \eqref{time2min2eq2}, $u_k(v_k(\eta))$ is a convex function.
\end{lemma}

\itshape {Proof:}  \upshape
We first prove that $v_k(\eta)$ is a convex function.
To show this, we define:
\vspace{-.5em}
\begin{equation}\label{apptime2min2eq1}\vspace{-.5em}
\phi(\eta)=\frac{s}{\eta},\quad 0\leq \eta\leq1,
\end{equation}
and
\vspace{-.5em}
\begin{equation}\label{apptime2min2eq2}\vspace{-.5em}
\varphi_k(\eta)={\frac {(1-\eta)T}a +\frac{A_k\log_2\eta}{f_k^{\max}}}, \quad 0\leq \eta\leq1.
\end{equation}
According to \eqref{time2min2eq3}, we have:
$
v_k(\eta)=\phi(\varphi_k(\eta)).
$
Then, the second-order derivative of $v_k(\eta)$ can be given by:
\vspace{-0.5em}
\begin{align}\label{apptime2min2eq3}\vspace{-0.5em}
v_k''(\eta)&=\phi''(\varphi_k(\eta))(\varphi_k'(\eta))^2+\phi'(\varphi_k(\eta))\varphi_k''(\eta).
\end{align}
According to \eqref{apptime2min2eq1} and \eqref{apptime2min2eq2}, we have:
\vspace{-.5em}
\begin{equation}\label{apptime2min2eq1_1}\vspace{-.5em}
\phi'(\eta)=-\frac{s}{\eta^2} \leq 0, \quad \phi''(\eta)=\frac{2s}{\eta^3} \geq 0,
\end{equation}
and
\vspace{-.5em}
\begin{equation}\label{apptime2min2eq2_1}\vspace{-.5em}
\varphi_k''(\eta)= -\frac{A_k}{(\ln2)f_k^{\max}\eta^2}  \leq 0.
\end{equation}
Combining \eqref{apptime2min2eq3}-\eqref{apptime2min2eq2_1}, we can find that $v_k''(\eta)\geq 0$, i.e., $v_k(\eta)$ is a convex function.

Then, we can show that $u_k(\eta)$ is an increasing and convex function.
According to the proof of Lemma 4, $u_k(\eta)$ is the inverse function of the right hand side of (\ref{time2min2}b).
If we further define function:
\vspace{-.5em}
\begin{equation}\label{apple2time2min2eq5_1}\vspace{-.5em}
z_k(\eta)=\eta\log_2\left(1+ \frac{g_kp_k^{\max}}{N_0\eta}
\right), \quad \eta \geq 0,
\end{equation}
$u_k(\eta)$ is the inverse function of $z_k(\eta)$, which gives
$u_k(z_k(\eta))=\eta$.

According to (\ref{apptime2min2eq5_1}), function $z_k(\eta)$ is an increasing and concave function, i.e., $z_k'(\eta)\geq 0$ and $z_k''(\eta)\leq 0$.
Since $z_k(\eta)$ is an increasing function, its inverse function $u_k(\eta)$ is also an increasing function.

Based on the definition of concave function, for any $\eta_1\geq0$, $\eta_2\geq 0$ and $0\leq\theta\leq1$, we have:
\vspace{-.5em}
\begin{equation}\label{apple2time2min2eq5_3}\vspace{-.5em}
z_k(\theta\eta_1+(1-\theta)\eta_2) \geq \theta z_k(\eta_1)+
(1-\theta)z_k(\eta_2).
\end{equation}
Applying the increasing  function $u_k(\eta)$ on both sides of \eqref{apple2time2min2eq5_3} yields:
\vspace{-.5em}
\begin{equation}\label{apple2time2min2eq5_5}\vspace{-.5em}
 \theta\eta_1+(1-\theta)\eta_2  \geq u_k(\theta z_k(\eta_1)+
(1-\theta)z_k(\eta_2)).
\end{equation}
Denote $\bar \eta_1=z_k(\eta_1)$ and $\bar \eta_2=z_k(\eta_2)$, i.e.,
we have $ \eta_1=u_k(\bar\eta_1)$ and $ \eta_2=u_k(\bar\eta_2)$.
Thus, \eqref{apple2time2min2eq5_5} can be rewritten as:
\begin{equation}\label{apple2time2min2eq5_6}
\theta u_k(\bar\eta_1)+(1-\theta)u_k(\bar\eta_1)  \geq u_k(\theta \bar\eta_1 +
(1-\theta) \bar \eta_2 ),
\end{equation}
which indicates that $u_k(\eta)$ is a convex function.
As a result, we have proven that $u_k(\eta)$ is an increasing and convex function, which shows:
\vspace{-.5em}
\begin{equation}\label{apple2time2min2eq5_7}\vspace{-.5em}
u_k'(\eta)\geq 0, \quad u_k''(\eta)\geq 0.
\end{equation}

To show the convexity of $u_k(v_k(\eta))$, we have:
\vspace{-.5em}
\begin{align}\label{apptime2min2eq6}\vspace{-.5em}
u_k''(v_k(\eta))=u_k''(v_k(\eta))(v_k'(\eta))^2+u_k'(v_k(\eta))v_k''(\eta) \geq 0,\nonumber
\end{align}
according to $v_k''(\eta)\geq 0$ and \eqref{apple2time2min2eq5_7}.
As a result, $u_k(v_k(\eta))$ is a convex function.  \hfill $\Box$

Lemma \ref{lemmatime2} implies that the optimization problem in \eqref{time2min2eq1} is a convex problem, which can be effectively solved. 
By finding the optimal solution of \eqref{time2min2eq1},  the sufficient and necessary condition for the feasibility of set (\ref{time2min2}a)-(\ref{time2min2}e) can be simplified
using the following theorem. 

\begin{theorem}\label{theoremtime1}
With a fixed $T$, set (\ref{time2min2}a)-(\ref{time2min2}e) is nonempty if and only if
\vspace{-0.5em}
\begin{equation}\label{time2min2eq5}\vspace{-0.5em}
 B\geq \sum_{k=1}^K u_k(v_k(\eta^*)),
\end{equation}
where $\eta^*$ is the solution to
$
 \sum_{k=1}^K   u_k'(v_k(\eta^*))v_k'(\eta^*)=0.
$
\end{theorem}

Theorem \ref{theoremtime1} directly follows  from Lemmas  \ref{lemmatime11} and \ref{lemmatime2}.
Due to the convexity of function $u_k(v_k(\eta))$, $ \sum_{k=1}^K   u_k'(v_k(\eta^*))v_k'(\eta^*)$ is an increasing function of $\eta^*$. As a result, the unique solution of $\eta^*$ to $ \sum_{k=1}^K   u_k'(v_k(\eta^*))v_k'(\eta^*)=0$ can be effectively solved via the bisection method.
\begin{algorithm}[t]
\caption{Delay Minimization}
\scriptsize
\begin{algorithmic}[1]
 \STATE Initialize $T_{\min}$, $T_{\max}$, and the tolerance $\epsilon_0$.
 \REPEAT
 \STATE Set $T=\frac{T_{\min}+T_{\max}}{2}$.
 \STATE Check the feasibility condition (\ref{time2min2eq5}).
 \STATE If set (\ref{time2min2}a)-(\ref{time2min2}e) has  a feasible solution, set $T_{\max}=T$. Otherwise, set $T_{\min}=T$.
 \UNTIL $(T_{\max}-T_{\min})/T_{\max}\leq \epsilon_0$.
\end{algorithmic}
\end{algorithm}
Based on Theorem \ref{theoremtime1}, the algorithm for obtaining the minimal delay is summarized in Algorithm 2.

\vspace{-.5em}
\section{Simulation Results}\vspace{-.5em}
For our simulations, we deploy $K=50$ users uniformly in a square area of size $500$ m $\times$ $500$~m with the BS located at its center.
The path loss model is $128.1+37.6\log_{10} d$ ($d$ is in km)
and the standard deviation of shadow fading is $8$ dB \cite{yang2020energyefficient}.
In addition, the noise power spectral density is  $N_0=-174$ dBm/Hz.
{We use the real open blog feedback dataset in \cite{buza2014feedback}.
This dataset with a total number of 60,021 data samples originates from blog posts and the dimensional of each data sample is 281.
The prediction task associated with the data is the prediction
of the number of comments in the upcoming 24 hours.}
Parameter $C_k$ is uniformly distributed in $[1,3]\times10^4$ cycles/sample.
{The effective switched capacitance in local computation is $\kappa=10^{-28}$.}
In Algorithm~1, we set  $\xi=1/10$, $\delta=1/10$, and $\epsilon_0=10^{-3}$. 
Unless specified otherwise, we choose an equal maximum average transmit power $p_1^{\max}=\cdots=p_K^{\max}=p^{\max}=10$ dBm, an equal maximum computation capacity $f_1^{\max}=\cdots=f_K^{\max}=f^{\max}=2$ GHz, a transmit data size $s=28.1$ kbits,
 and a bandwidth $B=20$ MHz.
 Each user has $D_k=500$ data samples, which are randomly selected from the dataset with equal probability.
All statistical results are averaged over 1000 independent runs.

\begin{figure}
\centering
\includegraphics[width=3.0in]{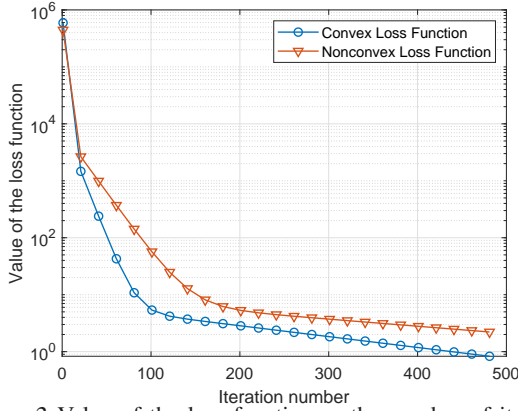}
\vspace{-1.5em}
\caption{Value of the loss function as the number of iterations varies for convex and nonconvex loss functions.}
\vspace{-2em}
\label{edfig1}
\end{figure}

In Fig. \ref{edfig1}, we show the value of the loss function as the number of iterations varies for convex and nonconvex loss functions.
For this feedback prediction problem, we consider two different loss functions:
convex loss function $f_1(\boldsymbol w,\boldsymbol x, y)=\frac1 2 (\boldsymbol x^T\boldsymbol w-y)^2$,
and
nonconvex loss function $f_2(\boldsymbol w,\boldsymbol x, y)=\frac1 2 (\max\{\boldsymbol x^T\boldsymbol w,0\}-y)^2$.
From this figure, we can see that, as the number of iterations increases, the value of the loss function first decreases rapidly and then decreases slowly for both convex and nonconvex loss functions.
According to Fig. \ref{edfig1}, the initial value of the loss function is $F(\boldsymbol w^{(0)})=10^6$ and the value of the loss function decreases to $F(\boldsymbol w^{(500)})=1$ for convex loss function after 500 iterations.
For our prediction problem, the optimal model $\boldsymbol w^{*}$ is the one that predicts the output without any error, i.e., the value of the loss function value should be $F(\boldsymbol w^{*})=0$.
Thus, the actual accuracy of the proposed algorithm is $\frac{F(\boldsymbol w^{(500)})-F(\boldsymbol w^{*})}{F(\boldsymbol w^{(0)})-F(\boldsymbol w^{*})}=10^{-6}$ after 500 iterations.
Meanwhile, Fig. \ref{edfig1}
clearly shows that the FL algorithm with a convex loss function can converge faster than that the one having a nonconvex loss function.
According to Fig. \ref{edfig1}, the loss function monotonically decreases as the number of iterations varies for even nonconvex loss function, which indicates that the proposed FL scheme can also be applied to the nonconvex loss function.

\begin{figure}[t]
\centering
\includegraphics[width=3.0in]{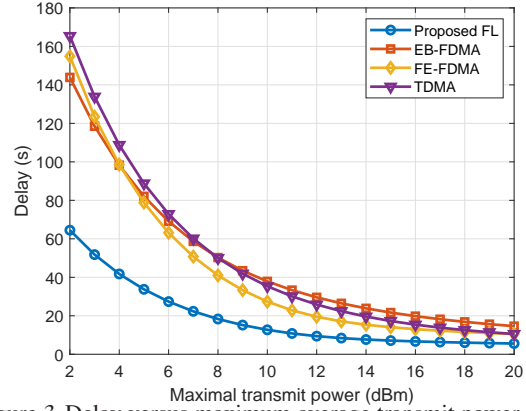}
\vspace{-1.5em}
\caption{Delay versus maximum average transmit power of each user.} \label{fig1}
\vspace{-2em}
\end{figure}

We compare the proposed FL scheme with the FL FDMA scheme with equal bandwidth $b_1=\cdots=b_K$ (labelled as `EB-FDMA'),
the FL FDMA scheme with fixed local accuracy $\eta=1/2$ (labelled as `FE-FDMA'), and the FL time division  multiple access (TDMA) scheme in \cite{tran2019federated} (labelled as `TDMA').
Fig. \ref{fig1} shows how the delay changes as the maximum average transmit power of each user varies.
We can see that the delay of all schemes decreases with the maximum average transmit power of each user.
This is because a large maximum average transmit power can decrease the transmission time between users and the BS.
We can clearly see that the proposed FL scheme achieves the best performance among all schemes.
This is because the proposed approach jointly optimizes bandwidth and local accuracy $\eta$, while the bandwidth is fixed in EB-FDMA and $\eta$ is not optimized in FE-FDMA.
Compared to TDMA, the proposed approach can reduce the delay by up to 27.3\%.

\vspace{-1em}
\section{Conclusions}
\vspace{-.5em}
{\color{myc1}{In this paper, we have investigated the delay minimization problem of FL over wireless communication networks.
The tradeoff between computation delay and transmission delay is determined by the learning accuracy.
To solve this problem, we first proved that the total delay is a convex function of the learning accuracy.
Then, we have obtained the optimal solution by using the bisection method.
Simulation results show the various properties of the proposed solution.}}

%


\bibliography{example_paper}
\bibliographystyle{icml2020}


\end{document}